\documentclass{article}
\usepackage{amsmath}  

\usepackage{PRIMEarxiv}

\usepackage[utf8]{inputenc} 
\usepackage[T1]{fontenc}    
\usepackage{hyperref}       
\usepackage{url}            
\usepackage{booktabs}       
\usepackage{amsfonts}       
\usepackage{nicefrac}       
\usepackage{microtype}      
\usepackage{lipsum}
\usepackage{fancyhdr}       
\usepackage{graphicx}       
\graphicspath{{media/}}     
\usepackage{tcolorbox} 
\usepackage{float} 

\title{Machine Assistant with Reliable Knowledge: Enhancing Student Learning via RAG-based Retrieval}

\author{
  Yongsheng Lian \\
  EduQuery AI \\
  Louisville, KY 40223\\
  \texttt{eduqueryai@gmail.com} \\
}

\begin{document}
\maketitle

\begin{abstract}

We present \textbf{Machine Assistant with Reliable Knowledge (MARK)}, a retrieval-augmented question-answering system designed to support student learning through accurate and contextually grounded responses. The system is built on a retrieval-augmented generation (RAG) framework, which integrates a curated knowledge base to ensure factual consistency. To enhance retrieval effectiveness across diverse question types, we implement a hybrid search strategy that combines dense vector similarity with sparse keyword-based retrieval. This dual-retrieval mechanism improves robustness for both general and domain-specific queries. The system includes a feedback loop in which students can rate responses and instructors can review and revise them. Instructor corrections are incorporated into the retrieval corpus, enabling adaptive refinement over time. The system was deployed in a classroom setting as a substitute for traditional office hours, where it successfully addressed a broad range of student queries. It was also used to provide technical support by integrating with a customer-specific knowledge base, demonstrating its ability to handle routine, context-sensitive tasks in applied domains. \textbf{MARK is publicly accessible at \href{https://app.eduquery.ai}{app.eduquery.ai}}.

\end{abstract}

\section{Introduction}

College education is facing significant challenges as increasing numbers of students arrive underprepared, particularly in foundational subjects like mathematics. Recent studies show that many university students struggle with basic mathematical concepts and lack the skills necessary for college-level coursework \cite{binkley2023math}. According to the National Assessment of Educational Progress (NAEP), nearly 40\% of high school graduates demonstrate insufficient proficiency in mathematics \cite{naep2022,sparks2024mathanxiety}. This academic gap hinders their performance in college and contributes to broader issues of retention and achievement. While one-on-one tutoring has proven effective, it is logistically impractical in large-enrollment courses. The scalability of traditional office hours is further constrained by scheduling conflicts among faculty, teaching assistants, and students—many of whom juggle academic responsibilities with part-time employment. These limitations reduce access to timely, personalized support when it is most needed.

Traditional question-answering (Q\&A) systems, such as expert systems, are capable of addressing student inquiries by retrieving the most relevant text excerpts from a predefined corpus. Because these responses are drawn from verified, educator-approved materials, they tend to be accurate and trustworthy~\cite{fernandez2009teaching,ott2010informationretrieval}. These systems often emphasize step-by-step problem-solving but lack the capacity to interpret or explain the underlying concepts behind a question. Their performance is highly dependent on the quality, structure, and coverage of the underlying knowledge base. Moreover, they are typically not well aligned with evolving curriculum objectives or the individualized needs of learners~\cite{henkel2024ragmath}.

Large language models (LLMs) have achieved impressive results across a wide spectrum of natural language tasks~\cite{brown2020fewshot,chowdhery2023palm,touvron2023llama,jaech2024openai,guo2025deepseek}. They achieve high scores on a wide range of college-level benchmarks like AP, GRE and SAT tests~\cite{achiam2023gpt4,shao2024deepseekmath}.  LLMs even reach the highest performance on challenging Olympiad math~\cite{trinh2024solving,chervonyi2025geometry}.   LLMs can be used in higher education. Most recent studies on the use of LLMs in higher education focus on high-level discussions of ethics, policy, and general applications, rather than exploring subject-specific implementations~\cite{rasul2023chatgpt_education,dempere2023impact,mvondo2023ethical}. While these studies acknowledge the potential of AI to enhance learning, they also highlight risks such as academic misconduct, bias, and misinformation, which can hinder the development of critical thinking skills. However, there is a limited research focused on integrating AI into specific courses—largely due to the scarcity of conversational educational data~\cite{dakshit2024rag}.

\subsection{Hallucination from LLMs}

Despite their remarkable capabilities, LLMs suffer from a critical limitation: hallucination-the generation of outputs that are coherent and plausible but factually incorrect or misleading~\cite{ji2023hallucination,kasneci2023chatgpt}. These inaccuracies are particularly problematic in educational contexts, where students may unknowingly adopt false information, leading to flawed conceptual understanding and persistent misconceptions~\cite{dziri2022faithdial}. The risk is amplified when AI-generated content is integrated into curricula or assessments without proper verification~\cite{zhui2024ethical}.

This challenge becomes even more pronounced in specialized domains such as aerospace engineering. For instance, a recent study evaluating LLMs in aerospace manufacturing revealed that the models frequently produced incorrect or misleading answers to domain-specific queries, posing a risk to students developing accurate mental models of critical technical concepts~\cite{liu2025llm_evaluation}.

Furthermore, students with limited subject matter expertise are especially vulnerable to overreliance on AI-generated responses. Without the skills to independently verify information, they may accept erroneous outputs at face value, compounding the problem of misinformation in learning environments~\cite{zhai2024effects}.

These hallucinations inherent from the fundamental architectural and training limitations. First,  LLMs are trained to predict the next word/token, not to verify the fact~\cite{ji2023hallucination}. They may generate plausible but incorrect information that fits learned patterns.  Second, the parametric nature of LLMs imposes an upper bound on their knowledge capacity~\cite{zhang2023retrieve}. OpenAI's GPT-4 is reportedly to have 1.8 Trillions of parameters, much larger than 85 billions of neurons in our brain. However, no model can fully encode all human knowledge, especially for rare or domain-specific facts. Third, LLMs are trained on static data, meaning their knowledge becomes outdated over time~\cite{achiam2023gpt4}. This issue is particularly pronounced in domains with highly specialized or long-tail content, such as flight dynamics or fluid mechanics, where domain-specific data may be scarce or poorly represented in public datasets. As a result, LLMs may struggle to provide accurate and contextually relevant responses in these areas. Last, LLMs are trained on vast internet-scale corpora, which contain factual inaccuracies, contradictions, and misinformation~\cite{shuster2021retrieval}. For example, when asking LLMs to explain lift on a wing, they still generated answers that the velocity on the top surface is higher than the bottom surface because air needs to reach the trailing edge at the same time even though the answer is incorrect and has been corrected by professors in the classroom.

\subsection{LLMs+RAG }

Retrieval-Augmented Generation (RAG) has emerged as a promising approach to mitigate some of the challenges faced by LLMs, particularly hallucination and limited knowledge generalization~\cite{lewis2020rag,bechard2024hallucination,kirchenbauer2024hallucination,niu2023ragtruth}. RAG is an architecture designed to enhance LLMs by integrating external knowledge retrieval into the text generation process.  The external knowledge typically comes from verified and curated sources, such as Wikipedia, technical manuals, textbooks, lecture notes, or proprietary knowledge bases. Since the retriever accesses content at inference time, RAG models can ground outputs in reliable documents, ensuring responses are both relevant and verifiable. Studies have shown that RAG-based systems can produce more factually accurate and specific outputs across a variety of tasks~\cite{shuster2021retrieval,gao2023rag_survey}. For instance, Lewis et al.~\cite{lewis2020rag} demonstrated that RAG significantly outperformed parametric-only models on several open-domain Q\&A benchmarks, while also enabling models to cite evidence and improve transparency.  Studies have shown that RAG offer a balance between adaptive responses and curriculum alignment~\cite{thus2024rag}. 

Additionally RAG facilitates knowledge freshness by retrieving from up-to-date corpora during inference. This capability enables LLMs to adapt to the most updated knowledge without costly retraining. RAG can also handle specialized tasks by searching custom sources like scientific papers without costly retraining of LLMs.

Several universities have successfully integrated LLM+RAG systems into college courses, demonstrating their efficiency and effectiveness in supporting student learning. At the University of Pennsylvania, ChemTAsk—a RAG-enabled open-source tool—was deployed in an advanced biological chemistry course to answer student questions using lecture notes and primary literature. The study showed ChemTAsk matched the performance of human teaching assistants. Students rated it as accurate, helpful, and faster than TAs~\cite{perez2025chemtask}. Similarly, the University of Michigan’s AI-University (AI-U) system was used in a graduate-level finite element methods course to deliver instructor-aligned responses by leveraging lecture videos, notes, and textbooks. The system demonstrated strong alignment with course materials~\cite{faghih2025aiuniversity}.  The University of California, Irvine used RAGMan, an AI-powered tutoring system designed for an introductory programming course~\cite{ma2024ai_tutors}. RAGMan utilizes RAG to provide students with homework and general programming assistance without giving direct answers. The system demonstrated high accuracy for questions within its intended scope. Student feedback indicated that the AI tutors were helpful for learning, with 78\% of users reporting improved learning outcomes~\cite{ma2024ai_tutors}. Feng et al.~\cite{feng2024courseassist} has used LLM-based tutoring system tailored for computer science education. Student feedback showed that the system increased the accessibility of course-specific tutoring help and shortened the feedback loop.

While LLM+RAG systems have shown promise in enhancing college-level instruction, outstanding challenges remain to limit their broader effectiveness and scalability.  First, studies have shown that the quality of user input material  directly affects the accuracy and relevance of generated responses~\cite{perez2025chemtask,faghih2025aiuniversity}. Without curated, comprehensive and unbiased content, these systems risk producing incomplete or misleading answers. Second, students may over-reliance on AI tools, potentially reducing engagement with core learning activities such as critical reading, discussion, or problem-solving~\cite{ma2024ai_tutors}. Third, hallucinated responses remain a challenge, particularly in open-ended or ambiguous queries~\cite{thway2024battling}. Last, many systems require substantial instructor effort to configure and align the AI responses with course pedagogy, which can be a barrier to adoption, especially in resource-limited settings.

We present MARK (Machine Assistant with Reliable Knowledge), a general-purpose LLM+RAG system designed to support both educational and technical domains by delivering accurate, contextually grounded responses. MARK is built on a RAG framework. To optimize retrieval across diverse query types, MARK employs a hybrid search mechanism that integrates both sparse and dense retrieval methods. A built-in feedback system enables students to rate responses and allows instructors to review and revise outputs. Beyond education, MARK can also be deployed for technical support.

\section{Retrieval Methods}

An effective RAG system relies critically on the quality of its passage retrieval component, which identifies and selects the most relevant context passages from a large corpus to support accurate response generation. To meet the demands of diverse and often complex user queries, the retrieval method must not only be computationally efficient, but also semantically precise, ensuring that the retrieved passages are both relevant and contextually appropriate.

\subsection{Keyword-Based Search}

Keyword-based search is a simple and widely used approach for information retrieval. With this search, documents are selected based on the presence of exact query terms. This method relies on direct matching of query keywords against document content, often using Boolean logic or exact string matching. While fast and interpretable, keyword search does not account for term importance or semantic meaning, making it less effective in handling synonyms, paraphrases, or ambiguous queries~\cite{schutze2008informationretrieval}.

\subsection{Sparse Vector Space Models}

Sparse vector space models represent documents and queries as high-dimensional vectors in which most elements are zero. They capture only the presence and frequency of specific words. Approaches such as Term Frequency–Inverse Document Frequency (TF-IDF)~\cite{ramos2003tfidf} and BM25~\cite{robertson2009bm25} exemplify this class of methods.

Like keyword-based methods, sparse vector space models rely on exact term matching. These models also rely on statistical weighting such as term frequency and inverse document frequency. These models assign higher weighting factors to terms that are more effective at differentiating between documents, which improves the accuracy of document ranking in response to a query. Despite their simplicity, these models remain popular due to their efficiency and strong performance in low-resource or domain-specific retrieval scenarios.

In these models, a collection of $N$ documents is represented as  $\mathcal{D} = \{d_i\}_{i=1}^N$,  and the vocabulary consists of $M$ unique terms as $\{t_j\}_{j=1}^M$. Each document $d_i$ is represented by a vector $\vec{d}_i \in \mathbb{R}^M$, where the $j$-th component of the vector is the weight of term $t_j$ in document $d_i$.

\subsubsection{TF-IDF}

In the  TF-IDF model, the weigh of term $t_j$ in a document $d_i$ is computed based on term frequency (TF), which measures how often $t_j$ appears in $d_i$,  and inverse document frequency (IDF), which measures how rare $t_j$ is acorss the entire document collection. The TF is defined as follows:

\begin{equation}
\text{TF}(t, d) = \frac{f_{t,d}}{\sum_{t' \in d} f_{t',d}},
\end{equation}

where $f_{t,d}$ is the number of times term $t$ appears in document $d$. The more frequently a term appears in a document compared to other terms, the higher its TF value.

The IDF  is defined as follows:

\begin{equation}
\text{IDF}(t, D) = \log \left( \frac{N}{|\{d \in D : t \in d\}|} \right),	
\end{equation}

where $N$ is the total number of documents and $|\{d \in D : t \in d\}|$ is the number of documents containing term $t$.  The less frequently a term appears across all documents in the corpus, the higher its IDF value. This reflects the idea that rare terms are more informative and carry greater discriminative power when identifying the content of a specific document.

Finally the TF-IDF Weight is calculated as the product of TF and IDF values as follows:

\begin{equation}
	\text{TF-IDF}(t, d, D) = \text{TF}(t, d) \cdot \text{IDF}(t, D)
	\label{eq:IDF}
\end{equation}

With TF-IDF, the whole corpus is represented by a matrix of shape $N \times M$. 

To identify the document that best matches a user query using TF-IDF, the query is first represented as a TF-IDF vector. The TFs are computed based on the terms represent in the query, while the IDFs  are derived from the full document corpus 
$D$. This ensures the query representation is consistent with the document vectors, allowing for accurate similarity comparisons. 

Next, we compute the cosine similarity between the query vector $\vec{q}$ and each document vector $\vec{d_i}$ in the corpus. Cosine similarity measures the angle between two vectors in high-dimensional space, capturing how closely aligned they are in terms of content:

\begin{equation} \label{eq:cosin}
	\text{similarity}(\vec{q}, \vec{d}_i) =  \frac{\vec{q} \cdot \vec{d}_i}{\|\vec{q}\| \cdot \|\vec{d}_i\|}
\end{equation}

Documents with the highest similarity scores are considered the most relevant and are selected as the retrieval results.

\subsubsection{Best Match 25(BM25)}
\label{sec:BM25}

BM25 is a ranking function used in information retrieval to estimate the relevance of documents to a given query~\cite{robertson2009bm25}. It is based on a bag-of-words model and extends the TF-IDF approach discussed above with more robust term frequency saturation and document length normalization.

For a multi-word query $Q = \{q_1, q_2, ..., q_m\}$, the BM25 score of document $d_j$ is:

\begin{equation}
\text{BM25}(Q, d_j) = \sum_{i=1}^{m} \text{IDF}(q_i)  \frac{f(q_i, d_j)  (k_1 + 1)}{f(q_i, d_j) + k_1  \left(1 - b + b \frac{|d_j|}{\text{avgdl}} \right)}
\end{equation}

Where:

- $q_i$: each term/word in the query.

- $f(q_i, d_j)$: frequency of term $q_i$ in document $d_j$, i.e., the raw count of how many times $q_i$ appears in document $d_j$.

- $|d_j|$: length of document $d_j$

- $\text{avgdl}$: average document length in the corpus

- $k_1$: term frequency scaling constant (typically ~1.5)

- $b$: length normalization constant (typically ~0.75)

The IDF used in BM25 is different from that used in TF-IDF (Eq.~\ref{eq:IDF}). The formula uses a log function that reduces the impact of very rare terms and stabilize scoring. The BM25 IDF is calculated as:

\begin{equation}
\text{IDF}(q_i) = \log \left( \frac{N - n(q_i) + 0.5}{n(q_i) + 0.5} + 1 \right)
\end{equation}

Where: 
- $N$: total number of documents.
- $n(q_i)$: number of documents that contain term $q_i$.

\subsection{Dense Vector Search }

Dense vector search, also known as semantic retrieval, has emerged as a foundational technique in modern information retrieval systems. It represents queries and documents as low-dimensional, dense vectors using deep learning models such as BERT or other transformer-based encoders~\cite{devlin2019bert}. Unlike sparse methods or keyword-based method that rely on exact word matches, dense retrieval captures the semantic meaning of text, enabling it to retrieve relevant passages even when there is no word match with the query~\cite{karpukhin2020dense}.  This paradigm is particularly effective for tasks requiring semantic matching, such as open-domain question answering, recommendation systems, and large-scale similarity search.

 In dense vector search 
 each document $d_i \in \mathcal{D}$ is mapped to a dense vector $\vec{v}_i \in \mathcal{R}^d$ via an embedding function:

\begin{equation}
	\vec{v}_i = f_d(d_i; \theta),
	\label{eq:vector_mapping}
\end{equation}

where $\mathcal{D} = \{d_i\}_{i=1}^N$ denotes a corpus of documents, $f_d: \mathcal{X} \to \mathcal{R}^d$ is typically parameterized by a deep neural network, and $\theta$ denotes the learnable parameters.

Given a query $q \in \mathcal{X}$, its vector representation is computed as:

\begin{equation}
	\vec{v}_q = f_q(q; \theta)
\end{equation}

The retrieval task is similar to the TF-IDF  by compute the similarity between the query vector $\vec{v_q}$ and the dense vector $\vec{v_i}$ as shown in Eq.\ref{eq:cosin}.

Previous studies have shown  when evaluated on a wide range of open-domain Q\&A datasets, dense retriever outperforms the BM25 system~\cite{karpukhin2020dense}. Recent advancements have made dense retrieval more practical and powerful, especially when combined with approximate nearest neighbor (ANN) techniques like FAISS~\cite{johnson2019similarity}.

The ANN makes dense retrieval more efficient. In ANN, the retrieval process seeks

\begin{equation}
	\arg\max_{\vec{v}_i \in \mathcal{V}} \text{sim}(\vec{v}_q, \vec{v}_i),
\end{equation}

where $\mathcal{V} = \{\vec{v}_i = f(d_i; \theta)\}_{i=1}^N$. However, exact nearest neighbor search in high-dimensional spaces is computationally expensive, especially when 
$N$ is large. ANN algorithms address this challenge by trading a small amount of accuracy for a substantial gain in speed and scalability.

Several ANN techniques have become standard in dense retrieval system.  Hierarchical Navigable Small World (HNSW) graphs (Malkov \& Yashunin, 2018) constructs a multi-layer graph structure that allows logarithmic-time search over the dataset. It provides high recall and fast retrieval even in large-scale applications. Product Quantization (PQ)~\cite{jegou2010product}  compresses vectors by dividing them into subvectors and quantizing each subvector independently. This enables efficient distance computation in compressed space and dramatically reduces memory footprint.

The retrieval efficiency can be further improved by using FAISS (Facebook AI Similarity Search)~\cite{johnson2019similarity}. FAISS is a widely adopted ANN library that supports various indexing strategies, including HNSW, PQ, and inverted file systems. It is optimized for both CPU and GPU execution, making it suitable for real-time, large-scale retrieval tasks.

Given a collection of $N$ dense vectors $\{\vec{v}_i \in \mathbb{R}^d\}_{i=1}^N$, FAISS supports construction of various index structures $\mathcal{I}$ to efficiently solve the problem:

\begin{equation}
	\vec{v}_{\text{NN}} = \arg\max_{\vec{v}_i \in \mathcal{V}} \text{sim}(\vec{v}_q, \vec{v}_i),	
\end{equation}

for a query vector $\vec{v}_q \in \mathbb{R}^d$, using similarity functions such as inner product or cosine similarity.   In our system, we adopt FAISS to enable fast and accurate hybrid retrieval.

\subsection{Hybrid search} 

Hybrid search leverages the complementary strengths of sparse and dense retrieval methods to enhance information retrieval, particularly in open-domain search and question-answering tasks. Sparse retrieval techniques, such as BM25 and TF-IDF, are effective at identifying exact term matches and highlighting rare but important keywords. In contrast, dense retrieval methods, which use neural embeddings, excel at capturing semantic similarity even in the absence of lexical overlap. By integrating both approaches—either through late fusion (e.g., combining retrieval scores) or unified indexing strategies—hybrid search systems can achieve higher recall and precision. Empirical studies demonstrate that hybrid retrieval consistently outperforms either method alone across a range of benchmarks~\cite{gao2021coil, sawarkar2024blended}.

\subsection{Re-Ranking}

In a RAG framework, re-ranking serves as a crucial intermediary between initial document retrieval and final answer generation by a LLM. The process begins by independently encoding the query $q$ and each document $d_i \in \mathcal{D}$ into vector representations using a shared or separate embedding function. These representations are then compared using a similarity function such as cosine similarity: 

\begin{equation}
	\text{sim}_{\text{retr}}(q, d_i) = \cos\left(f_q(q), f_d(d_i)\right),
\end{equation}

where $f_q$ and $f_d$ map the query and docment to their respective vector embeddings. When using the same encoder for both, the embeddings lie in a shared semantic space, improving alignment and enabling fast approximate retrieval. However, this bi-encoder setup encodes the query and document independently, which limits its ability to capture fine-grained semantic or syntactic interactions between the two.

To improve the quality of retrieved contexts, a re-ranking stage is applied using a more powerful cross-encoder model $s(q, d_i)$, which takes the query–passage pair as joint input:

\begin{equation}
	s(q, d_i) = f_{\text{cross}}([q; d_i]),
\end{equation}

where $[q; d_i]$ denotes the concatenation of the query $q$ and passage $d_i$, and $f_{\text{cross}}$ is a neural scoring model (e.g., BERTA~\cite{devlin2019bert}, RoBERTa~\cite{liu2019roberta} or SBERT~\cite{reimers2019sentencebert}) trained to predict relevance or entailment.

The re-ranked list $\{d_1', d_2', \ldots, d_k'\}$ is obtained by sorting the original candidates based on the cross-encoder scores $s(q, d_i)$:

\begin{equation}
	\text{ReRank}(q, \{d_i\}) = \text{argsort}_{i=1}^k \; s(q, d_i).
\end{equation}
These top-ranked passages are then passed to the generator, which uses them to produce the final response. Re-ranking is especially effective in domains where precision and grounding are critical, as it prioritizes passages that are not just semantically similar but also contextually and pedagogically aligned with the user's intent.

\section{Frontend Development}

This section describes the frontend architecture of the system, which manages all user interactions. The frontend is built using HTML and JavaScript, providing a responsive, accessible, and platform-independent user experience. It includes two main interfaces: a student interface, which allows users to submit questions, receive AI-generated responses, and provide feedback; and an instructor interface, which enables faculty to monitor student interactions, review and revise AI responses, and contribute additional curated content. The entire web interface is deployed on Google Cloud Platform (GCP) and delivered securely through Cloudflare, ensuring global availability, low latency, and protection against DDoS attacks.

\subsection{Student Interface}
\begin{figure}[htbp]
	\centering
	\includegraphics[width=0.6\textwidth]{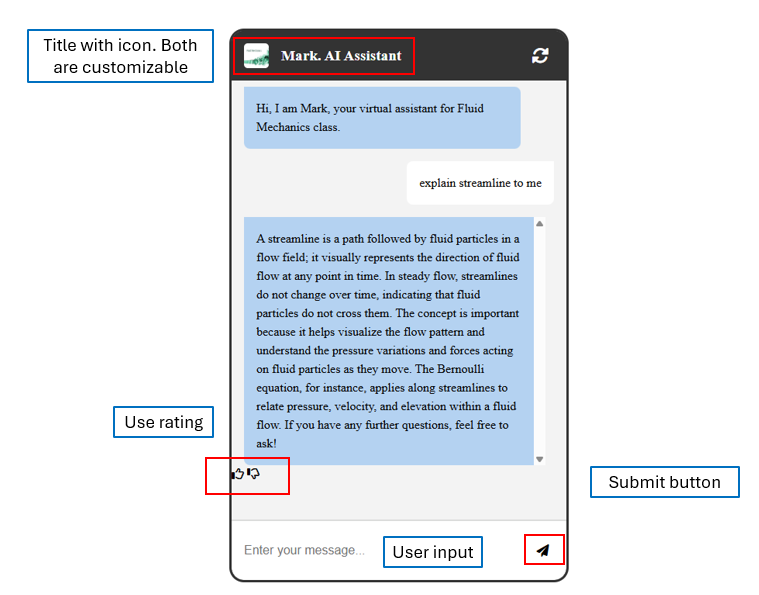} 
	\caption{Interface of MARK with Text Entry and User Feedback}
	\label{fig:user_ui}
\end{figure}

The student interface allows students or customers to submit natural language queries. As shown in Figure~\ref{fig:user_ui}, this interface includes a customizable header—featuring an icon and styling options that instructors or administrators can configure.

Users enter their questions into a text input field and submit them using a designated button. Upon submission, the frontend issues a secure API request to the backend service. The backend processes the query, performs retrieval and generation operations, and returns a response via an encrypted connection. The frontend then renders the result within the interface, providing a seamless question-answering experience.

\subsection{Instructor Interface}

\begin{figure}[htbp]
	\centering
	\includegraphics[width=0.6\textwidth]{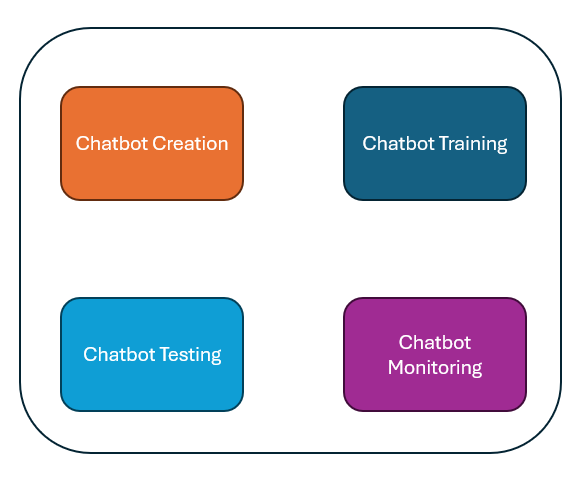} 
	\caption{The four frontend modules of the instructor interface}
	\label{fig:frontend_modules}
\end{figure}

The instructor interface is designed for instructors and IT support personnel to manage knowledge base content, oversee chatbot behavior, and provide corrective feedback. This interface offers four core functionalities: creating, training, testing, and monitoring chatbots, which are shown in Figure~\ref{fig:frontend_modules}.

\subparagraph{Chatbot Creation Panel}

\begin{figure}[htbp]
 	\centering
 	\includegraphics[width=0.6\textwidth]{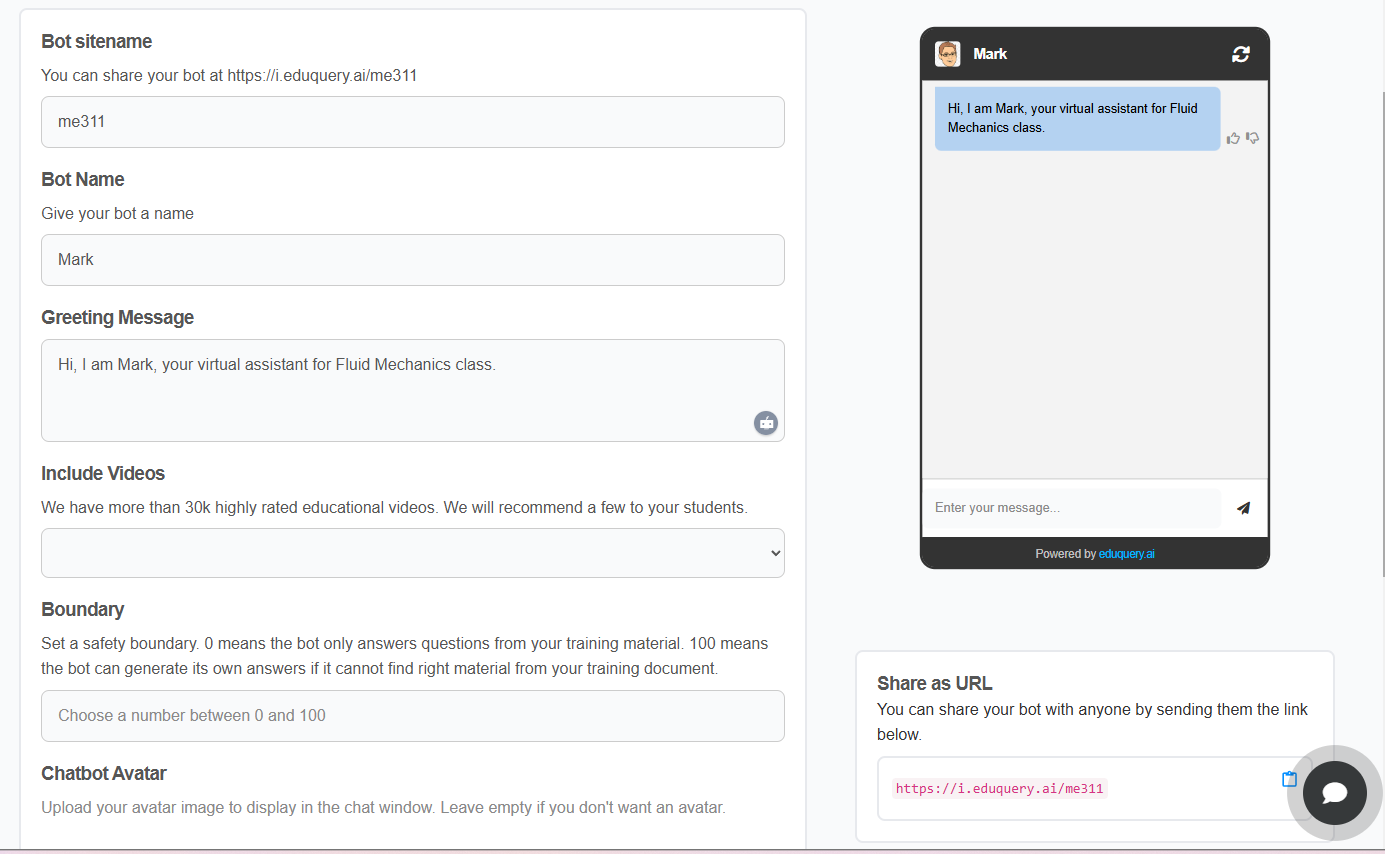} 
 	\caption{Interface of Create a New Chatbot}
 	\label{fig:create_bot}
 \end{figure}

Figure~\ref{fig:create_bot} displays the interface for initializing a new chatbot. In this view, users can assign a unique name to the chatbot and specify a custom greeting message to be presented during initial interactions. An additional configuration option allows users to control the degree to which the underlying LLM can incorporate external knowledge beyond the RAG database. This is represented as a tunable parameter: a value of 0 strictly limits the LLM to information retrieved from the RAG database, while a value of 100 allows unrestricted use of external knowledge sources by the LLM.

Once configured, the chatbot can be deployed either as a standalone HTML page or seamlessly embedded within an existing website, such as a popup widget or inline element.

 \subparagraph{Chatbot Training Panel}
 
\begin{figure}[htbp]
 	\centering
 	\includegraphics[width=0.6\textwidth]{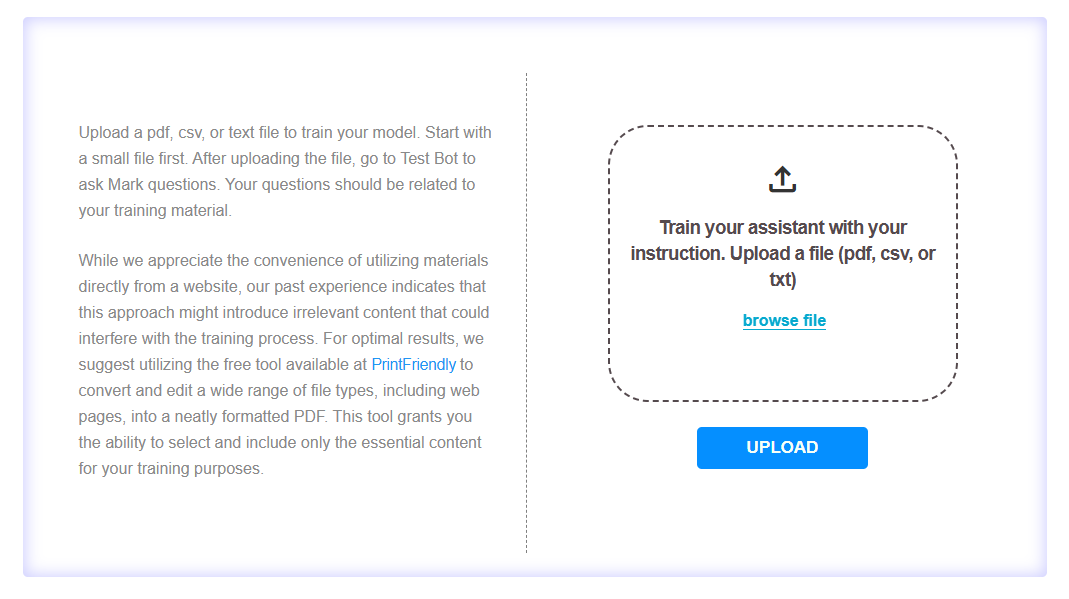} 
 	\caption{Interface of Uploading Contents for RAG}
 	\label{fig:train_bot}
 \end{figure}

Figure~\ref{fig:train_bot} shows the panel to upload documents into the RAG database. This panel supports uploading files in PDF, CSV, or plain text formats. For users who wish to incorporate content from a webpage, we recommend first converting the webpage into a PDF file to ensure consistent formatting and easier processing. Once uploaded, the documents undergo a series of preprocessing steps—such as text extraction, cleaning, and formatting—to ensure they are properly indexed for retrieval. The detailed procedure for integrating these documents into the RAG database will be addressed in the backend development section.

\subparagraph{Chatbot Testing Panel}
 
Following content ingestion, the chatbot can be evaluated through the Testing Panel, which is designed to simulate real-world user interactions. This interface mirrors the layout and functionality of the student interface UI shown in Figure~\ref{fig:user_ui}. This is to ensure consistency across testing and deployment environments. The Testing Panel enables instructors or system administrators to validate the chatbot's responses against expected outputs, assess performance accuracy, and identify potential failure cases prior to public deployment.

 \subparagraph{Chatbot Monitoring Panel}
 
\begin{figure}[htbp]
 	\centering
 	\includegraphics[width=0.6\textwidth]{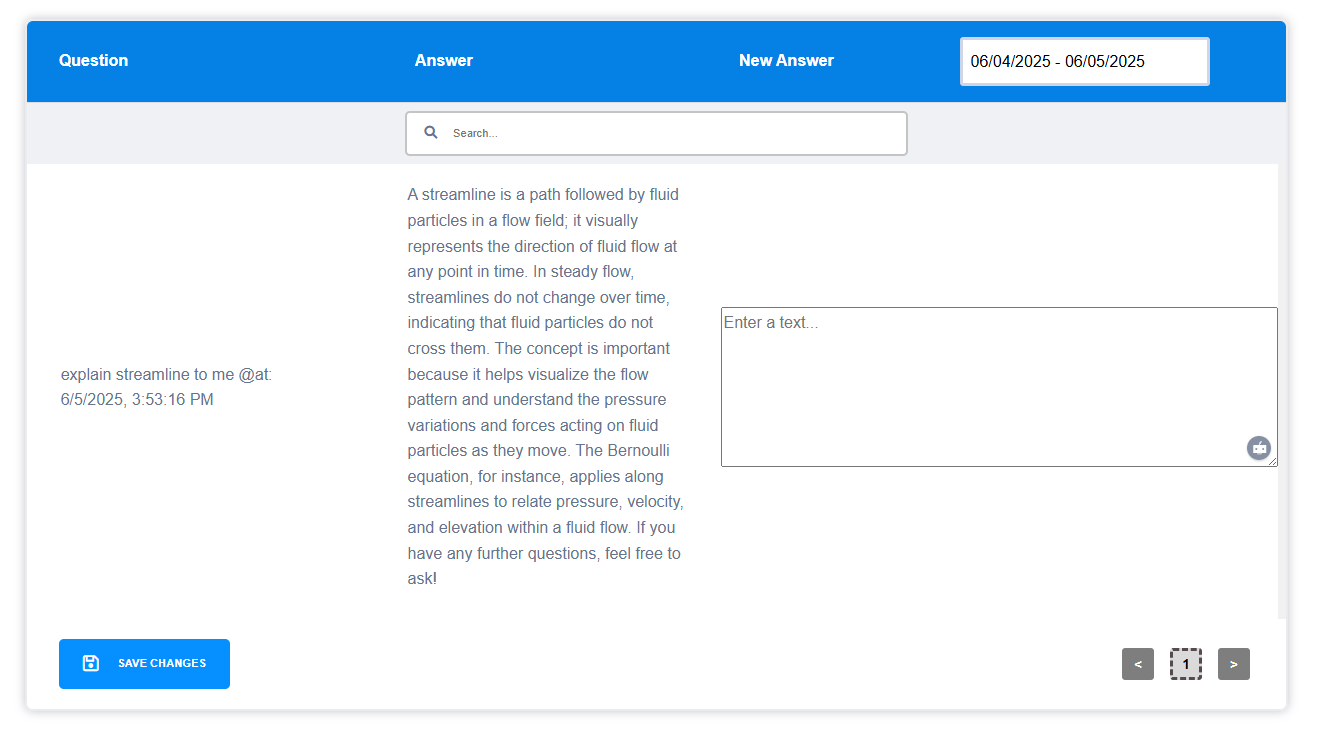} 
 	\caption{Interface of Monitoring AI Responses}
 	\label{fig:monitor_bot}
 \end{figure}

Instructors can review the chatbot’s responses and provide corrective feedback through the Monitoring Panel, as illustrated in Figure~\ref{fig:monitor_bot}. The panel allows users to filter and select chatbot responses generated within a specified time range. Revised or improved responses can be entered into the correction input area on the right. Upon clicking the Save button, the updated content is immediately inserted into the RAG  database, making it available for future inference queries.
 
 \section{Backend Development}
 
\begin{figure}[htbp]
 	\centering
 	\includegraphics[width=0.6\textwidth]{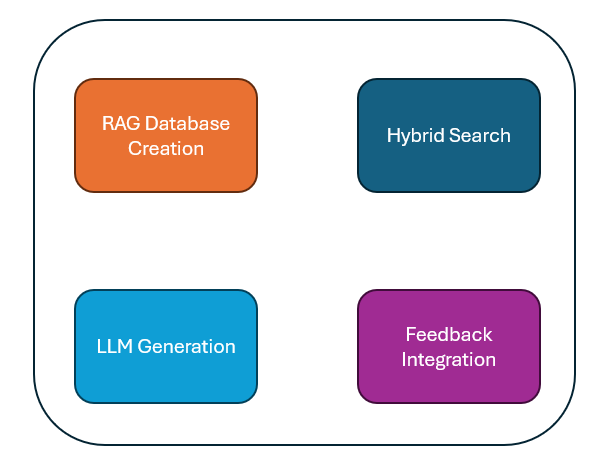} 
 	\caption{The four backend modules}
 	\label{fig:backend_modules}
 \end{figure}

 The backend architecture of MARK is implemented using the Python Flask web framework. Flask is a lightweight and efficient foundation for handling HTTP requests and serving the application’s API endpoints. This backend is responsible for orchestrating tasks such as chatbot training, RAG database updates, and response generation. The Flask server is deployed on Google Cloud. The four key modules in the backend are shown in Figure~\ref{fig:backend_modules}. 
 
 \subsection{RAG Database Preparation}

The uploaded documents from the frontend are processed through a backend API to prepare them for embedding and retrieval. For plain text and comma-separated values files, the content is segmented directly into smaller chunks. For PDF files, a preprocessing step is required to extract and convert the content into structured textual data. We developed a Python-based pipeline for this task using the PyMuPDF library, which allows direct access to each page and its associated text content.

The extraction process iterates over each page of the document, retrieving raw text data and optionally applying standard preprocessing techniques such as line break removal and whitespace normalization. The resulting cleaned text is then divided into semantically coherent chunks. Based on our experiments, including the section heading within each chunk enhances retrieval performance by providing additional context.

Chunk length should be tailored to the expected length of the response; shorter responses benefit from smaller chunks. For chatbot application, the answer length is usually short. Hence, we used a maximum chunk length of 384 tokens.

Each text chunk is subsequently embedded into dense vector representation, as described in  Eq.~\ref{eq:vector_mapping}. In general, higher-dimensional embeddings can capture more nuanced semantic information by providing a larger representational space. This is especially beneficial in large-scale retrieval systems where fine-grained distinctions between semantically similar documents are critical. However, for smaller corpora, the marginal gains from increasing the embedding dimension beyond a certain point tend to diminish, while the computational and memory costs grow linearly with dimensionality.

Our empirical results indicate that a 512-dimensional vector offers a good balance between performance and efficiency for our use case. In our application, we employed the `jina-embeddings-v2-small-en` model~\cite{günther2023jina}, a monolingual English model with a maximum input length of 8192 tokens. Based on a BERT-like architecture, this model is pretrained with sequences of up to 512 tokens but can generalize to longer sequences. The embedding tool map each chunk into a 512-dimension dense vector. With only 33 million parameters, it offers highly efficient and fast inference while maintaining competitive performance for RAG applications. For multi-language embedding, jina3~\cite{günther2025jinaembeddingsv4universalembeddingsmultimodal} can be used.

\subsection{Hybrid RAG Retrieval}

When a user query is submitted, the backend initiates a hybrid retrieval process that combines both sparse and dense search paradigms to maximize relevance and coverage. For the sparse retrieval component, we employ the BM25 algorithm (see Section~\ref{sec:BM25}). BM25 can identify documents containing exact keyword matches.

In parallel, dense retrieval is performed using vector similarity search powered by FAISS. Each text chunk in the RAG database is pre-encoded into a fixed-size dense embedding vector using the Jina Embeddings v2 model. The query is embedded into the same vector space, and FAISS retrieves the top-ranked chunks based on vector similarity (see Equation~\ref{eq:cosin}). We then merge the top results from the BM25 and FAISS queries into a unified context document. This combined context is then re-ranked and the top $k$ passages are forwarded to LLM for final response generation.

\subsection{LLM and its Prompt}

The retrieved document is subsequently passed to a LLM to generate a response. For this purpose, we utilize the OpenAI API. To ensure accurate and contextually relevant answers, we provide a carefully designed prompt that instructs the LLM to ground its responses strictly in the retrieved content. The prompt reads as follows:

\begin{tcolorbox}[colback=gray!5, colframe=gray!40, title=MARK Prompt, fonttitle=\bfseries ,coltitle=black]

\begin{tcolorbox}[colback=red!5, colframe=red!40, sharp corners, boxrule=0.5pt, left=2pt, right=2pt, top=2pt, bottom=2pt]
You are a knowledgeable and patient teaching assistant. Refer to the previous conversation when answering, and politely decline if a question is outside your knowledge. Be concise.
\end{tcolorbox}

\end{tcolorbox}

This prompt encourages the model to adopt a helpful and precise tone, while constraining its output to information supported by the retrieved document, thereby reducing the likelihood of hallucinations and improving factual consistency.

\subsection{Integrate Feedback into RAG Database}

An essential component of the MARK is its feedback integration mechanism. It enables users to review and refine the AI-generated responses. Corrected or improved answers can be submitted directly through the instructor interface. Once received, this feedback is incorporated into the existing RAG  knowledge base. On the backend, the new text entries are first appended to the raw document repository. Subsequently, both the sparse index  and the dense vector index  are updated to reflect the new content. For small to moderately sized datasets—fewer than one million entries—index regeneration is computationally efficient. In our deployment, these updates are executed on a Google Cloud instance equipped with 4 Intel Cascade Lake CPUs and 32 GB of RAM.

\section{Results}

\subsection{MARK as Virtual Tutor}

We developed a Fluid Mechanics module using MARK for Fluid Mechanics class. Fluid Mechanics is a required course in Mechanical Engineering field. It covers fundamental topics like hydrostatics, Bernoulli's equation, control volume analysis, dimensional analysis, and internal and external flows. We created the basic RAG database using class syllabus, class notes, and homework problems. The database is further enhanced with curated answers to student questions by instructors.  The experiment was conduced in the fall of 2023.

Students accessed MARK via a dedicated link provided by the instructor. Students can submit questions in natural language at any time. Students often asked MARK about administrative and logistical aspects of the course, such as exam schedules, the scope of material covered in upcoming exams, homework deadlines, and clarification of class policies. Although these details were clearly documented in the syllabus, the frequency of such queries suggests that students preferred a quick, centralized, and easily accessible source of answers over traditional course documentation.

In addition to logistical questions, students also posed academic queries, particularly related to homework assignments. These questions were typically brief, such as “How to solve Homework 5.6,” reflecting a desire for direct assistance. MARK responded with complete solutions rather than offering step-by-step guidance. This pattern suggests the potential value of refining MARK’s instructional strategies through example prompts or usage training to encourage deeper learning.

Students also posed playful questions such as 'Mark, who is your dad?', reflecting the natural curiosity that often accompanies the early adoption of emerging technologies like generative AI. These interactions suggest that students were not only testing the bot’s capabilities, but also relating to it as a conversational partner—much like how they might engage with a personality on a social media app.

Student inputs to MARK are limited to text only, without the ability to upload diagrams, equations, or their intermediate work. This limits MARK's ability to assess student understanding or to provide feedback tailored to their specific problem-solving approach. Future iterations may benefit from incorporating multimodal input or answer checking features to better support formative assessment.

Overall adoption during the 2023 pilot phase was limited, mirroring national hesitations and skepticism following the widespread attention sparked by the release of ChatGPT-3.5 in November 2022. This hesitation was further compounded by the absence of clear institutional policies or guidance on the appropriate use of AI tools in the classroom.

\subsection{MARK as Technical Support}

MARK can be tailored to function as an intelligent technical support assistant, designed to address a wide range of customer inquiries. Unlike traditional FAQ pages, which require users to manually scan through lengthy and often irrelevant lists of answers, or rule-based chatbots that rely on rigid decision trees and canned responses, MARK delivers dynamic, context-aware support. It interprets user queries in natural language, retrieves relevant knowledge from an external document corpus, and generates human-like responses in real time. This allows MARK to handle a far broader and more nuanced range of questions than systems constrained by static templates or pre-scripted flows. By offering immediate, 24/7 assistance tailored to individual customer needs, MARK significantly improves the efficiency and effectiveness of technical support while reducing user frustration and support latency.

A key advantage of MARK is its ability to draw on vast knowledge bases—including product documentation, troubleshooting guides, and historical support records—to generate accurate, context-aware responses in real time. Since many customer queries are repetitive or follow recognizable patterns, the system can retrieve similar past cases or relevant technical documents and synthesize them into coherent, actionable responses using its generative capabilities. This allows MARK to deliver more helpful and comprehensive answers than traditional rule-based systems, while continuously improving as new data becomes available.

We show a few examples to demonstrate what MARK can do in technical support and customer service. 

\begin{tcolorbox}[colback=gray!5, colframe=gray!40, title=Give Instructions, fonttitle=\bfseries ,coltitle=black]

\begin{tcolorbox}[colback=red!5, colframe=red!40, sharp corners, boxrule=0.5pt, left=2pt, right=2pt, top=2pt, bottom=2pt]
\textbf{Customer:} I need help setup my wifi router
\end{tcolorbox}

\vspace{4pt}

\begin{tcolorbox}[colback=blue!5, colframe=blue!40, sharp corners, boxrule=0.5pt, left=2pt, right=2pt, top=2pt, bottom=2pt]
\textbf{MARK:} To set up your wifi router, visit \textcolor{orange}{m.******.com}.  
If you need help, email us at \textcolor{orange}{contact@****.com}.  
\end{tcolorbox}

\begin{tcolorbox}[colback=red!5, colframe=red!40, sharp corners, boxrule=0.5pt, left=2pt, right=2pt, top=2pt, bottom=2pt]
\textbf{Customer:} I would like to change my plan
\end{tcolorbox}

\vspace{4pt}

\begin{tcolorbox}[colback=blue!5, colframe=blue!40, sharp corners, boxrule=0.5pt, left=2pt, right=2pt, top=2pt, bottom=2pt]
\textbf{MARK:} To change your plan, please log in to your account and navigate to the 'Add Data' page. From there, you can select a new plan. Any remaining data from your current plan will roll over, ensuring you don't lose any unused data. If you encounter any issues, feel free to ask for assistance!

\end{tcolorbox}
\end{tcolorbox}

\begin{tcolorbox}[colback=gray!5, colframe=gray!40, title=Troubleshoot, coltitle=black , fonttitle=\bfseries]

\begin{tcolorbox}[colback=red!5, colframe=red!40, sharp corners, boxrule=0.5pt, left=2pt, right=2pt, top=2pt, bottom=2pt]
\textbf{Customer:} my wifi hotspot does not work. 
\end{tcolorbox}

\vspace{4pt}

\begin{tcolorbox}[colback=blue!5, colframe=blue!40, sharp corners, boxrule=0.5pt, left=2pt, right=2pt, top=2pt, bottom=2pt]
\textbf{MARK:}  If your device does not work, please follow these troubleshooting steps:

1. Ensure that your SIM card is activated.

2. Change your device's Access Point Name (APN) settings.

3. Restart your device after making the APN change.

4. Confirm that your device is an unlocked 4G device.

If you still encounter issues, email us at \textcolor{orange}{contact@****.com} for further assistance.

\end{tcolorbox}
\end{tcolorbox}

MARK can be integrated with internal databases and customer relationship management systems, which enables MARK to provide highly personalized assistance, such as verifying customer email addresses, checking account balances, or diagnosing issues with a specific device registered to a particular user. This multi-source integration enables MARK not only to understand general technical knowledge but also to act on individualized customer data. This personalized, context-aware support significantly reduces the need for human intervention in routine inquiries, increases operational efficiency, and enhances overall customer satisfaction.

\begin{tcolorbox}[colback=gray!5, colframe=gray!40, title=Function calls, coltitle=black , fonttitle=\bfseries]

\begin{tcolorbox}[colback=red!5, colframe=red!40, sharp corners, boxrule=0.5pt, left=2pt, right=2pt, top=2pt, bottom=2pt]
\textbf{Customer:} Sim card not working 
\end{tcolorbox}

\vspace{4pt}

\begin{tcolorbox}[colback=blue!5, colframe=blue!40, sharp corners, boxrule=0.5pt, left=2pt, right=2pt, top=2pt, bottom=2pt]
\textbf{MARK:} To assist you with your SIM card not working, please provide your SIM card's ICCID number (which should be 19-20 digits long). This will help me check its status and determine if there are any issues such as expiration or lack of data.

\end{tcolorbox}

\begin{tcolorbox}[colback=red!5, colframe=red!40, sharp corners, boxrule=0.5pt, left=2pt, right=2pt, top=2pt, bottom=2pt]
\textbf{Customer:} \textcolor{orange}{*********24107496794}
\end{tcolorbox}

\vspace{4pt}

\begin{tcolorbox}[colback=blue!5, colframe=blue!40, sharp corners, boxrule=0.5pt, left=2pt, right=2pt, top=2pt, bottom=2pt]
\textbf{MARK:} The SIM card with ICCID \textcolor{orange}{*********24107496794} is active and currently has a plan of \textcolor{orange}{*****} at a cost of \textcolor{orange}{***}. It will not expire until June 27, 2026. If you're experiencing issues, please check the device settings or let me know how I can further assist you!
\end{tcolorbox}

\end{tcolorbox}

\section{Conclusion}

The development and initial testing of MARK demonstrate the potential of RAG-based AI systems to address key challenges in education and technical support. By integrating carefully curated knowledge with LLMs, MARK is able to generate more accurate and trustworthy responses, reducing the risk of hallucination.

Another key benefit of MARK is its ability to support many students without requiring more effort from instructors. The system can handle common student questions, guide them with helpful prompts, and improve over time by learning from user ratings and instructor feedback. These features make MARK an effective tool for both classroom education—such as in engineering courses—and technical support tasks for customers. Its success shows the value of human feedback.

However, there is still room for improvement. MARK depends heavily on the quality and depth of its knowledge base, which means experts must keep updating and reviewing the content. Also, the current system only works with text, limiting its usefulness in subjects like engineering that often rely on images, equations, and diagrams. Future versions of MARK should include the ability to process different kinds of information (multimodal data) and offer more personalized feedback based on each student’s needs. 

Our testing also shows that the AI system cannot fully replace human support. While MARK is effective in handling common and repetitive queries, many customers still prefer or specifically request interaction with a real person.  Some customers can also quickly recognize when they are interacting with an AI chatbot, showing the gap between machine-driven and human support. Such limitations emphasize the need for continued improvement in AI conversational abilities.

\bibliographystyle{unsrt}
\bibliography{Mark_RAG_LLN}

\end{document}